\documentclass{article}

\usepackage{PRIMEarxiv}

\usepackage[utf8]{inputenc} 
\usepackage[T1]{fontenc}    
\usepackage{hyperref}       
\usepackage{url}            
\usepackage{booktabs}       
\usepackage{amsfonts}       
\usepackage{nicefrac}       
\usepackage{microtype}      
\usepackage{lipsum}
\usepackage{fancyhdr}       
\usepackage{graphicx}       
\graphicspath{{media/}}     

\usepackage{multirow}
\usepackage{array}
\newcolumntype{P}[1]{>{\centering\arraybackslash}p{#1}}
\newcolumntype{M}[1]{>{\centering\arraybackslash}m{#1}}
\usepackage{graphicx}
\usepackage{amsmath}
\usepackage{amssymb}
\usepackage{float}
\usepackage{tcolorbox}
\usepackage{xcolor}
\usepackage{lineno}

\title{Topology-Adaptive Seismic Response Prediction of Interconnected Structural Systems Using Reduced Order Model Embedded Graph Neural Network
}

\author{
  Huangbin Liang\\
  Department of Civil Environmental \\ and Geomatic Engineering \\
  ETH Zürich \\
  Zurich, Switzerland\\
  \texttt{huangbin.liang@ibk.baug.ethz.ch}
   \And
  Beatriz Moya \\
  ENSAM Institute of Technology \\
  Paris, France\\
  Corresponding email:\\
  \texttt{beatriz.moya\_garcia@ensam.eu}\\
  \\
  \AND
  Francisco Chinesta\\
  ENSAM Institute of Technology \\
  Paris, France\\
  CNRS@CREATE LTD., CNRS \\
  Singapore\\
  \And
  Eleni Chatzi \\
  Department of Civil Environmental \\ and Geomatic Engineering \\
  ETH Zürich \\
  Zurich, Switzerland\\
}

\begin{document}
\maketitle

\begin{abstract}
Electrical substation equipment is commonly interconnected by busbars and flexible conductors, whose mechanical interactions can significantly influence seismic response. Such interconnected infrastructures exhibit complex dynamic interactions, where the response of each component is influenced by local structural properties, mechanical coupling, system topology, and transient external excitation. High-fidelity finite element models can capture some of these effects, but their computational cost limits their use for real-time prediction and repeated evaluation in digital twin applications. Purely data-driven models offer faster inference, yet they typically struggle to generalize when the system layout, component combination, or connection properties change, requiring additional simulations and retraining for each new configuration. 
Motivated by the paradigm of physics-enhanced machine learning, this work presents a reduced-order model (ROM) embedded Graph Neural Network (ROMeGNN) for topology-aware seismic digital twins of interconnected structural systems. The system is represented as a graph, where structural components are treated as nodes and mechanical interconnections as edges. Each high-fidelity component model is first reduced to a reduced-order modal representation, which forms a differentiable dynamics operator within the graph neural network. Rather than learning the full response directly, the graph neural network estimates topology-dependent interaction forces that enter the differentiable ROM solver as coupling inputs for modal state advancement. This hybrid formulation preserves the known component-level physics and uses data-driven message passing to learn complex inter-component forces across varying system layouts, enabling topology generalization and real-time prediction.
The framework is demonstrated on interconnected electrical substation equipment subjected to randomized seismic inputs. It delivers rapid predictions of response quantities relevant to seismic damage assessment while remaining applicable across varying layouts, component combinations, and excitation scenarios. These results highlight the potential of embedding physics within graph learning architectures for real-time, interpretable, topology-adaptive seismic digital twins of interconnected engineering systems.

\end{abstract}


\keywords{physics enhanced machine learning \and digital twins \and  Graph neural networks \and reduced-order model \and interconnected structural systems.}
\section{Introduction}
Electric power systems play a crucial role in modern society, providing the backbone for essential functions ranging from residential utility to industrial production. Among the key elements of these systems, electrical substations stand out as vital infrastructure to adjust power voltage and control power flow, ensuring efficient power transformation and distribution to end users across various geographic areas \cite{liang2021system}. However, substation electrical equipment is particularly vulnerable to seismic excitation because of the widespread use of brittle porcelain materials, low inherent damping, and slender structural configurations \cite{fujisaki2014seismic,zhu2024towards}. Damage to such equipment can interrupt power delivery, trigger cascading failures, and cause substantial economic and societal losses \cite{guo2017critical}.

\begin{figure}[htbp]
    \centering
    \includegraphics[width=0.85\linewidth]{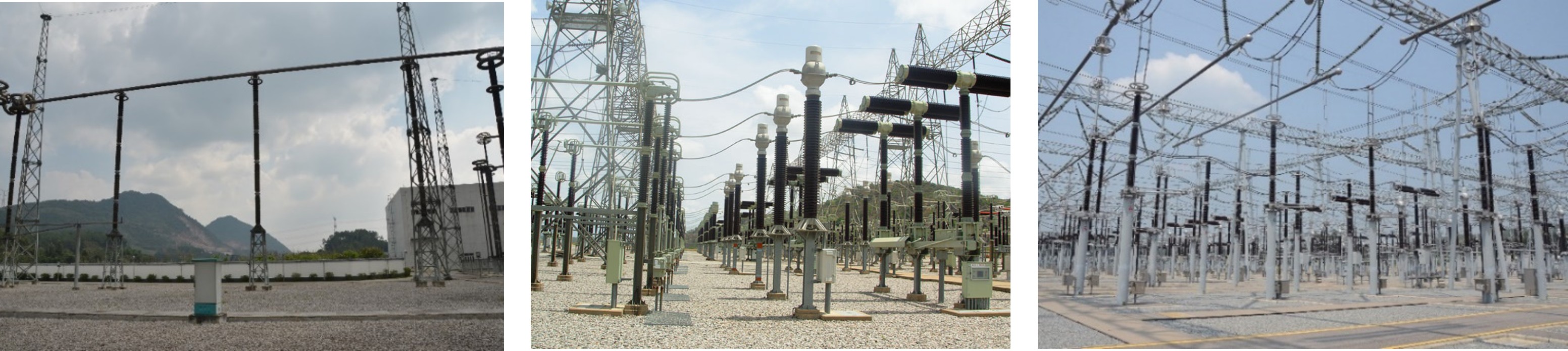}
    \caption{Electrical equipment interconnected by rigid busbars or flexible conductors within an electrical substation.}
    \label{substation-interconnected-equipment}
\end{figure}

Given the critical importance of substations to power-system reliability, extensive experimental studies have been conducted to evaluate and improve the seismic performance of substation electrical equipment \cite{xie2019seismic,gokcce2019seismic,xie2021seismic,liu2023seismic}. Shaking-table tests are widely used because they provide direct and reliable evidence of equipment dynamic behaviour, failure mechanisms, and seismic capacity. However, due to limitations in table size, payload capacity, and testing cost, most tests have focused on individual equipment \cite{gokcce2019seismic,xie2021seismic,liu2023seismic} or, at most, three connected components \cite{mohammadi2013investigation}. This individual component-level focus does not fully represent practical substations, where electrical equipment rarely behaves as isolated units. Adjacent devices are commonly interconnected through rigid busbars or flexible conductors, as illustrated in Figure \ref{substation-interconnected-equipment}. These interconnections can transmit interaction forces during earthquakes, causing the response of each component to depend not only on its own structural properties but also on neighboring equipment, connection characteristics, system topology, and transient ground motion. Field investigations and experimental studies have shown that such connection-induced interaction forces may contribute significantly to seismic damage, especially under strong excitation \cite{mohammadi2013investigation,xie2019influence}.

High-fidelity finite element (FE) modeling remains a powerful tool for analyzing the seismic behavior of interconnected power infrastructures \cite{guo2025seismic,zhu2024post,liang2025reliability}. Such models can represent geometric details and component flexibility, and compute local response quantities (e.g., stress at the bottom cross-section of the equipment) relevant to damage assessment. However, their computational cost becomes prohibitive when repeated simulations are required for uncertainty quantification, digital twin updating, seismic retrofitting optimization, or real-time scenario evaluation. This challenge is further amplified for interconnected electrical equipment, where the interaction dynamics are difficult to model explicitly. Rigid busbars with sliding end connections \cite{he2022seismic} and flexible conductors attached through clamp fittings \cite{dastous2007guidelines,xie2019influence} may introduce nonlinear coupling effects, localized sliding behavior, uncertain boundary conditions, and installation-dependent connection properties. In practice, these coupling mechanisms are rarely known with sufficient accuracy, and representing them using ideal spring-dampers or rigid links may introduce modeling errors, limiting purely simulation-based approaches in both efficiency and predictive reliability.

Reduced-order models (ROMs) provide an alternative route for efficient physics-based simulation. By projecting high-fidelity structural dynamics onto a low-dimensional modal or reduced basis, ROMs can substantially reduce computational costs while preserving essential structural dynamics \cite{padula2024brief,chinesta2011short,xie2024analytical}. Projection-based model reduction has been widely studied for large-scale dynamical systems and many-query simulations \cite{frohlich2022uncertainty,vlachas2024parametric}. In SHM and digital twin contexts, ROMs have been used to accelerate simulation-based training, online inference, and structural state estimation \cite{agathos2024accelerating,hollins2026projection,vlachas2024reduced}. Nevertheless, conventional ROMs are usually constructed for a prescribed structural system or a fixed set of parameters. When the system topology changes, such as when different equipment components are connected in different layouts, the reduced operators and coupling terms may need to be reassembled or redefined. Moreover, if inter-component forces are nonlinear or only partially known, a purely physics-based ROM may still require explicit constitutive assumptions for the coupling behavior.

Structural health monitoring (SHM) and data-driven modeling have provided new opportunities for rapid condition assessment and response prediction of engineering structures \cite{farrar2007introduction}. With the increasing availability of sensor data, simulation data, and data science tools, machine learning methods have been widely explored for damage detection, response reconstruction, and structural state estimation \cite{bao2021machine}. Classical SHM frameworks rely on measured response features and statistical pattern recognition to infer structural condition \cite{zhang2023rapid, xu2022real}, while recent deep learning methods have extended these ideas to high-dimensional time-series data, image-based inspection, and vibration-based diagnostics \cite{cha2024deep,mohammadkhorasani2023augmented,cunha2023review}. For post-earthquake assessment of substation equipment, however, directly instrumenting every device is often impractical because of cost, insulation constraints, and electromagnetic interference. In contrast, ground motion records are easier to monitor at the substation level and can be shared across equipment. This makes ground-motion-driven response prediction an attractive alternative for rapid assessment of interconnected electrical equipment \cite{zhu2024post,moya2025post}. Despite these advances, many data-driven models are trained for a fixed structural configuration and learn a direct mapping from excitation or sensor measurements to response or damage indicators \cite{zhu2024post,zhang2023rapid,xu2022real}. Such models may perform well within the training distribution but often struggle when the system topology, number of components, component types, or connection properties change. For topology-adaptive digital twins, this limitation is critical because retraining a separate model for every possible configuration would require extremely high computational costs and defeat the purpose of rapid deployment.

Graph neural networks (GNNs) offer a promising way to overcome this limitation by representing a structural system as a graph rather than as a fixed-size input vector. In this representation, components can be treated as nodes, and their mechanical or physical interconnections as edges. Through message passing, GNNs learn local interaction rules between connected components and aggregate this information over the system topology \cite{wu2020comprehensive}. As a result, the same learned model can be applied to systems with different numbers of components, connection layouts, and interaction patterns. This paradigm has made GNNs attractive for learning across applications ranging from particle interactions \cite{sanchez2020learning, moya2019learning}, and molecular dynamics \cite{park2021accurate}, to mesh-based simulations \cite{pfaff2020learning}, graph-structured fault diagnosis problems \cite{zhang2024graph}, and populations' behavior of structural elements\cite{duthe2024flexible,jian2025using}. In particular in structural seismic simulation contexts, Kuo et al. \cite{kuo2024gnn} have also demonstrated that graph embeddings can encode structural geometry and dynamic properties to predict nonlinear response histories across buildings with different configurations within a prescribed structural family. However, existing applications have primarily focused on individual structural systems and have not explicitly addressed physically interconnected systems composed of multiple standalone components with mechanical connections. In addition, without explicit enforcement of the governing equations, a purely data-driven GNN may generate physically inconsistent responses, accumulate errors during long rollouts, and require large amounts of training data to cover the relevant range of dynamic behaviors. 

\begin{figure}[htbp]
    \centering
    \includegraphics[width=0.8\linewidth]{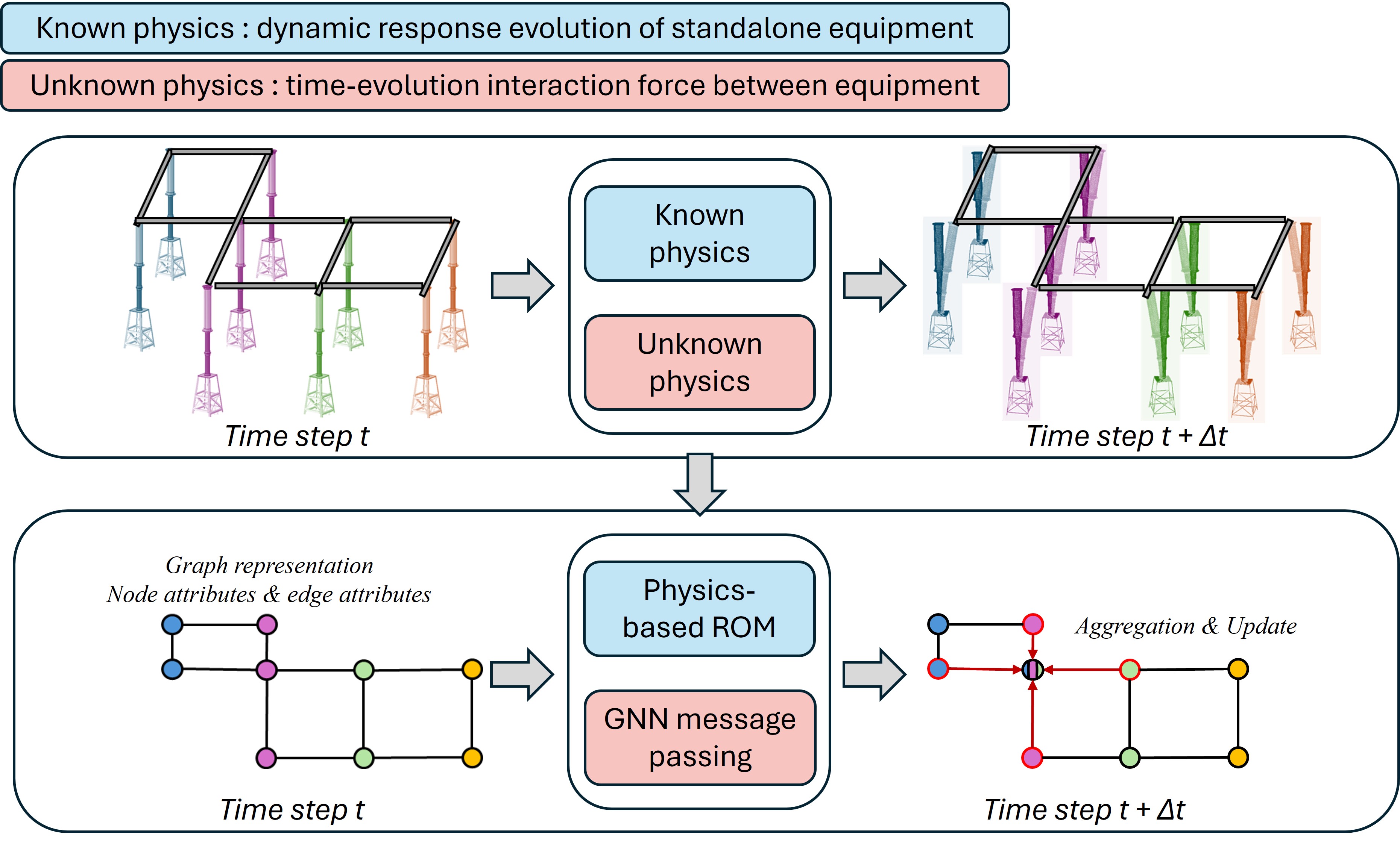}
    \caption{Concept of the proposed ROMeGNN forward simulator for interconnected electrical equipment. The known physics is the dynamic evolution of standalone equipment, represented by physics-based reduced-order models. The unknown physics is the interaction dynamics between connected equipment, learned by graph neural network message passing.}
    \label{physics-enhanced-forward-simulator}
\end{figure}

These above-mentioned limitations indicate that a suitable model for interconnected electrical equipment should be able to represent spatial relationships among components, capture temporal evolution under seismic excitation, share learned interaction rules across variable-size systems, and accommodate physical effects that are difficult to prescribe explicitly. In the present problem, the physics can be naturally separated into two parts: the known dynamic evolution of standalone equipment and the unknown interaction dynamics between connected equipment. Motivated by the paradigm of physics-enhanced machine learning \cite{liang2025harnessing,haywood2024discussing}, this study proposes a Reduced-Order-Model-embedded Graph Neural Network (ROMeGNN), in which physics-based ROMs preserve the component-level structural dynamics and computation efficiency, while a GNN learns the topology-dependent interaction effects that are not explicitly known. The concept of this hybrid forward simulator is illustrated in Figure \ref{physics-enhanced-forward-simulator}. The proposed model properly handles spatially located equipment and their interactions as nodes and edges in a graph. Temporal evolution is captured through autoregressive state advancement using the differentiable ROM solvers. In this way, component-level ROMs serve as differentiable dynamics solvers during time-domain rollout prediction, while the GNN provides the flexibility needed to handle variable system topologies and complex inter-component interactions.

The main contributions of this work are summarized as follows:
\begin{itemize}
    \item A physics-enhanced graph learning framework is developed for topology-adaptive seismic response prediction of interconnected electrical equipment systems,  where equipment components are represented as nodes and mechanical interconnections as edges.
    \item A differentiable ROM solver is embedded into the learning architecture to preserve the known component-level structural dynamics and improve long-term rollout consistency, reducing the burden of learning the entire dynamical system from data alone.
    \item A GNN-based interaction model is formulated to learn effective topology-dependent interaction forces between connected components, avoiding the need for explicit cable, contact, or boundary interaction models.
    \item The framework provides rapid predictions of both response quantities and physically meaningful seismic demand quantities, including displacement, velocity, interaction force, and section moment, for seen and unseen system configurations.
    \item The proposed method is evaluated and compared with baseline models under varying component combinations, connection properties, excitation scenarios, nonlinear coupling effects, and unseen system sizes, demonstrating its potential for real-time topology-adaptive digital twin applications.
\end{itemize}

\section{Methodology}
\label{sec:methodology}
Graphs neural networks serve as a means to model unstructured domains that range from mesh discretizations to populations. Any GNN architecture can be derived from the fundamental basis of geometric deep learning
\cite{bronstein2021geometric}. Meshgraphnets \cite{pfaff2020learning} represent a popular architecture in computational mechanics. However, Battaglia  et al. \cite{battaglia2018relational}  state the possibility of investigating different architectures to define graph neural networks based on relational reasoning and inductive biases. In a nutshell, the goal is encode available knowledge in the structure itself to adapt to the problem to be learnt. As the authors claim, GNNs have high flexibility in the definition of their structure as long as they respect the principles of permutation invariance and relational information present in the problem. Within this context, we propose embedding structural ROMs, which inherit the permutation-independent properties of finite element assembly, into the graph learning architecture. This allows established structural mechanics knowledge to be incorporated directly as a physics-based inductive bias.

\subsection{Overview of the proposed ROMeGNN}
\begin{figure}[htbp]
    \centering
    \includegraphics[width=1\linewidth]{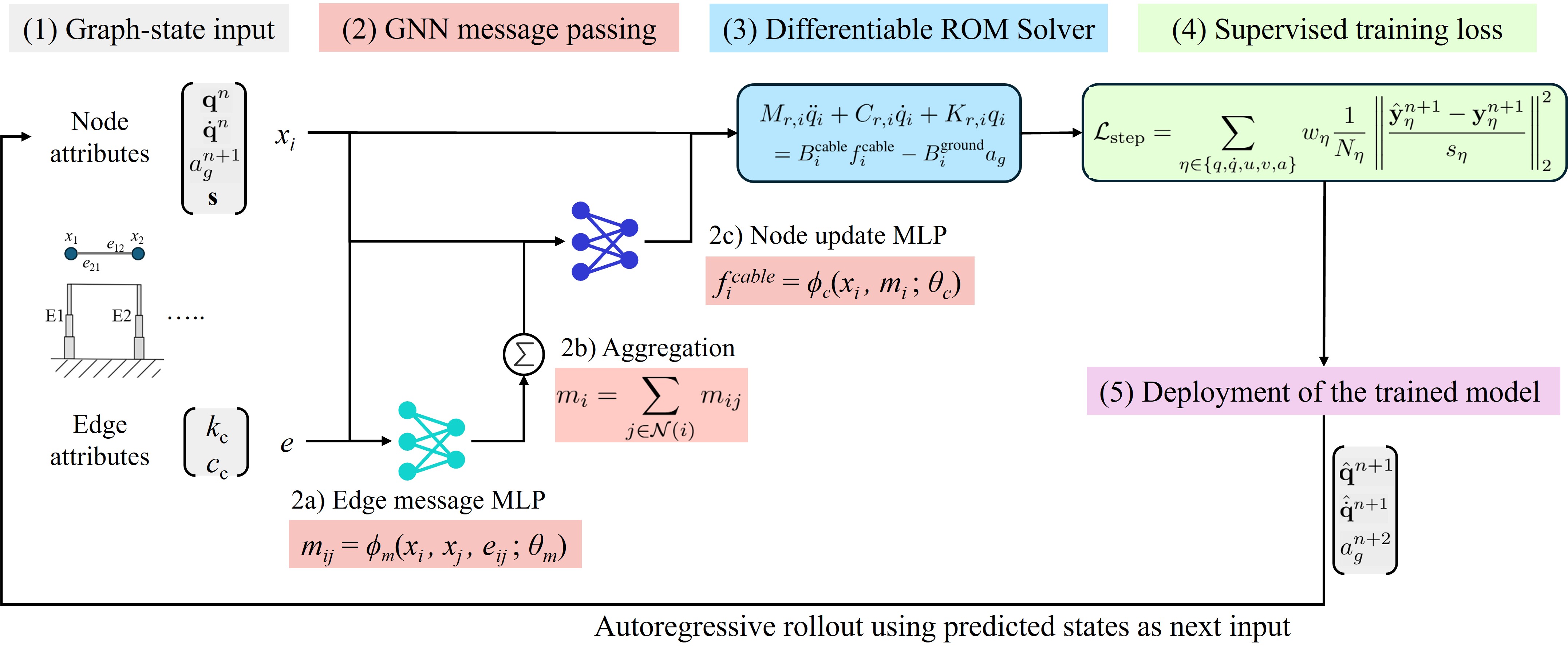}
    \caption{Overview of the training and deployment workflow for the proposed ROMeGNN. At each time step, modal states, ground acceleration, component and edge attributes are encoded as graph-state inputs. The message-passing GNN estimates the effective interaction force acting on each component, which is then passed to the differentiable ROM solver for physics-based state advancement. During training, the one-step prediction is supervised by ground truth from FOM reference states. During deployment, the predicted modal states are recursively fed back, together with the next ground-motion input, to perform autoregressive rollout prediction.}
    \label{ROM-Decoder-PEGNN}
\end{figure}

The proposed framework aims to predict the seismic response of interconnected electrical equipment systems with varying component combinations, connection properties, excitation
scenarios, and graph topologies. As illustrated in Figure \ref{ROM-Decoder-PEGNN}, the method combines a GNN with differentiable ROM solvers. The central idea is to separate the system dynamics into two parts: the known component-level structural dynamics and the unknown interaction dynamics between connected equipment. The former is represented by physics-based ROMs, while the latter is learned from data through graph message passing.

At each time step, the interconnected equipment system is represented as a graph $\mathcal{G}=(\mathcal{V},\mathcal{E})$, where each node denotes one equipment component and each edge denotes a mechanical interconnection. The graph-state input contains the current modal states of each component, the next-step ground acceleration, component-level attributes, and edge properties. Based on these inputs, the message-passing GNN estimates the effective topology-dependent interaction force acting on each component. Instead of directly predicting the next structural response, the learned interaction forces are passed into the ROM solver, which advances the modal states according to the reduced structural dynamic equations.

The model is trained as a one-step supervised solver using reference data generated from full-order model (FOM) simulations. During training, the predicted next-step modal states and reconstructed top responses are compared with the corresponding FOM reference quantities. During deployment, the trained model is used autoregressively: the predicted modal states at the current step are fed back as part of the graph-state input for the next step, together with the next ground-motion value. This enables time-domain rollout prediction using only the graph configuration, component attributes, connection properties, initial states, and ground-motion history.

\subsection{Component-level finite element model and modal ROM construction}
\label{sec:FEmodal_ROM_construction}
Before describing each block in Figure \ref{ROM-Decoder-PEGNN}, the component-level ROM construction is first introduced because it provides the physical operators used by the differentiable ROM solver. For each standalone equipment component, a high-fidelity finite element model is first constructed to describe its linear structural dynamics. The finite element model provides the mass, damping, and stiffness matrices, denoted by $\mathbf{M}_i$, $\mathbf{C}_i$, and $\mathbf{K}_i$ for the $i$-th component. Under horizontal base excitation, the dynamic equilibrium equation of an isolated component with an external interaction force applied at the top/interface degree of freedom (referred to as the top DOF hereafter) can be written as:
\begin{equation}
\mathbf{M}_i\ddot{\mathbf{u}}_i(t)
+
\mathbf{C}_i\dot{\mathbf{u}}_i(t)
+
\mathbf{K}_i\mathbf{u}_i(t)
=
-\mathbf{M}_i\mathbf{r}_i a_g(t)
+
\mathbf{b}_i f_i^{\mathrm{cable}}(t),
\label{eq:component_fom}
\end{equation}
where $\mathbf{u}_i(t)$ is the physical displacement vector, $\mathbf{r}_i$ is the influence vector associated with the base excitation direction, $a_g(t)$ is the ground acceleration, $\mathbf{b}_i$ maps the scalar interaction force to the top DOF, and $f_i^{\mathrm{cable}}(t)$ denotes the interaction force transmitted from connected equipment that is not explicitly known.

To obtain a compact representation of the component dynamics, modal reduction is applied to each standalone equipment model. The displacement field is approximated by a truncated modal basis:
\begin{equation}
\mathbf{u}_i(t)
\approx
\mathbf{\Phi}_i\mathbf{q}_i(t),
\label{eq:modal_expansion}
\end{equation}
where $\mathbf{\Phi}_i\in\mathbb{R}^{n_i\times r}$ contains the selected $r$ mode shapes and $\mathbf{q}_i(t)\in\mathbb{R}^{r}$ denotes the modal coordinate vector. The modal basis is obtained from the generalized eigenvalue problem:
\begin{equation}
\mathbf{K}_i\mathbf{\Phi}_i =
\mathbf{M}_i\mathbf{\Phi}_i\mathbf{\Omega}_i^2 ,
\label{eq:eigenproblem}
\end{equation}
where $\mathbf{\Omega}_i$ contains the retained natural circular frequencies.

Substituting Eq.~\eqref{eq:modal_expansion} into Eq.~\eqref{eq:component_fom} and premultiplying by $\mathbf{\Phi}_i^{T}$ gives the reduced-order dynamic equation:
\begin{equation}
\mathbf{M}_{r,i}\ddot{\mathbf{q}}_i(t)+\mathbf{C}_{r,i}\dot{\mathbf{q}}_i(t)+\mathbf{K}_{r,i}\mathbf{q}_i(t)
=
-
\mathbf{B}_{i}^{\mathrm{ground}} a_g(t)+\mathbf{B}_{i}^{\mathrm{cable}} f_i^{\mathrm{cable}}(t),
\label{eq:component_rom}
\end{equation}
where
\begin{equation}
\mathbf{M}_{r,i}=\mathbf{\Phi}_i^T\mathbf{M}_i\mathbf{\Phi}_i,\quad
\mathbf{C}_{r,i}=\mathbf{\Phi}_i^T\mathbf{C}_i\mathbf{\Phi}_i,\quad
\mathbf{K}_{r,i}=\mathbf{\Phi}_i^T\mathbf{K}_i\mathbf{\Phi}_i,
\label{eq:reduced_matrices}
\end{equation}
and
\begin{equation}
\mathbf{B}_{i}^{\mathrm{ground}}
=\mathbf{\Phi}_i^T\mathbf{M}_i\mathbf{r}_i,\quad
\mathbf{B}_{i}^{\mathrm{cable}}
=\mathbf{\Phi}_i^T\mathbf{b}_i .
\label{eq:input_mappings}
\end{equation}
Here, $\mathbf{B}_{i}^{\mathrm{ground}}$ maps the ground acceleration into the modal coordinates, while $\mathbf{B}_{i}^{\mathrm{cable}}$ maps the interaction force at the top DOF into the reduced-order model.

The displacement, velocity, and acceleration at the top DOF of each equipment component, all relative to the ground, are reconstructed from the modal state as:
\begin{equation}
u_{\mathrm{top},i}(t)
=\mathbf{H}_i\mathbf{q}_i(t),
\quad
v_{\mathrm{top},i}(t)
=\mathbf{H}_i\dot{\mathbf{q}}_i(t),
\quad
a_{\mathrm{top},i}(t)
=\mathbf{H}_i\ddot{\mathbf{q}}_i(t),
\label{eq:top_response_reconstruction}
\end{equation}
where $\mathbf{H}_i$ is the row vector extracted from the modal basis at the predefined top DOF. These reduced matrices and response-reconstruction operators are later used by differentiable ROM solvers in the proposed graph learning framework.

\subsection{Graph-state representation}
\label{sec:graph_state_representation}
The interconnected equipment system is represented as a graph
$\mathcal{G}=(\mathcal{V},\mathcal{E})$, where each node
$v_i\in\mathcal{V}$ denotes one equipment component and each edge
$e_{ij}\in\mathcal{E}$ denotes a mechanical interconnection between two components. The graph topology can vary with the number, type, and arrangement of equipment components, allowing the same model architecture to be applied to systems with different sizes and connection layouts.

At time step $n$, the dynamic state and static descriptors of each component are encoded into a node feature vector:
\begin{equation}
\mathbf{x}_i^n
=
\left[
\mathbf{q}_i^n,\,
\dot{\mathbf{q}}_i^n,\,
a_g^{n+1},\,
\mathbf{s}_i
\right],
\label{eq:node_feature}
\end{equation}
where $\mathbf{q}_i^n$ and $\dot{\mathbf{q}}_i^n$ are the modal displacement and velocity of component $i$, $a_g^{n+1}$ is the ground acceleration used for the one-step advancement from $n$ to $n+1$, and $\mathbf{s}_i$ denotes component-level static attributes. In this study, $\mathbf{s}_i$ includes the component type indicator and the retained modal frequencies.

For each edge, the edge attribute vector is defined as:
\begin{equation}
\mathbf{e}_{ij}
=
\left[
k_{c,ij},\,
c_{c,ij}
\right],
\label{eq:edge_feature}
\end{equation}
where $k_{c,ij}$ and $c_{c,ij}$ denote the connection stiffness and damping descriptors between components $i$ and $j$, respectively. The relative separation between connected components is implicitly encoded through the connection stiffness $k_{c,ij}$, which depends on the connection length. For each undirected physical connection, two directed graph edges are introduced with shared edge attributes, allowing bidirectional message passing between connected components.

In implementation, the dynamic quantities in $\mathbf{x}_i^n$ and the edge descriptors in $\mathbf{e}_{ij}$ are normalized before being passed to the GNN, and the normalization details are given in Section~\ref{sec:implementation_details}, while the ROM solver itself operates on physical-scale modal states and ground acceleration.

\subsection{GNN message passing for interaction-force learning}
The graph neural network is used to estimate the effective topology-dependent interaction force acting on each component. Instead of directly predicting the next structural response, the GNN learns an effective interaction force, which is later passed into the differentiable ROM solver for physics-based state advancement.

For each directed edge from node $j$ to node $i$, an edge message is computed as:
\begin{equation}
\mathbf{m}_{ij}
=
\phi_m
\left(
\mathbf{x}_i^n,
\mathbf{x}_j^n,
\mathbf{e}_{ij};
\theta_m
\right),
\label{eq:edge_message}
\end{equation}
where $\phi_m$ is an edge message multilayer perceptron (MLP) with trainable parameters $\theta_m$. The incoming messages are aggregated at each node using a permutation-invariant summation:
\begin{equation}
\mathbf{m}_{i}
=
\sum_{j\in\mathcal{N}(i)}
\mathbf{m}_{ij},
\label{eq:message_aggregation}
\end{equation}
where $\mathcal{N}(i)$ denotes the set of neighboring nodes connected to node $i$.

The normalized interaction force acting on component $i$ is then predicted by a node update MLP:
\begin{equation}
\hat{f}_{i}^{\mathrm{cable},n+1}
=
\phi_c
\left(
\mathbf{x}_i^n,
\mathbf{m}_{i};
\theta_c
\right),
\label{eq:force_prediction}
\end{equation}
where $\phi_c$ has trainable parameters $\theta_c$. Before being supplied to the ROM solver, the predicted cable interaction force is converted back to its physical scale and is applied at the top DOF of component $i$ during the state advancement from $t_n$ to $t_{n+1}$. Only a single message-passing round is performed during each state transition, with no additional graph propagation layers stacked within the same time step. This choice follows the division of roles in ROMeGNN: the ROM resolves the internal dynamics of each component, while the GNN estimates the interaction forces transmitted through direct physical connections. With the bidirectional edge construction introduced above, one-hop aggregation already incorporates information from all directly connected neighbors into the current force prediction, and longer-range dynamic dependencies are captured over successive rollout steps through the updated component states. This maintains a direct correspondence between the learned messages and the physical connections. Interactions beyond immediate neighbors are propagated through successive rollout steps.

The message and node update functions are shared across all edges and nodes, enabling the model to learn transferable local interaction rules and operate on graphs with unseen numbers of components or connection layouts.

\subsection{Differentiable ROM solver and response reconstruction}
\label{sec:Differentiable ROM decoder}
The differentiable ROM solver implements the reduced-order dynamics in Eq. \eqref{eq:component_rom} as a one-step state-advancement operator. At time step $n$, the GNN-predicted interaction force $\hat{f}_{i}^{\mathrm{cable},n+1}$ replaces the effective interaction force in the ROM, while the ground acceleration $a_g^{n+1}$ is used as the known external excitation. Given the current modal states $\mathbf{q}_i^n$ and $\dot{\mathbf{q}}_i^n$, the solver advances the state of component $i$ as:
\begin{equation}
\left(
\hat{\mathbf{q}}_i^{n+1},
\hat{\dot{\mathbf{q}}}_i^{n+1},
\hat{\ddot{\mathbf{q}}}_i^{n+1}
\right)
=
\mathcal{D}_{\mathrm{ROM}}
\left(
\mathbf{q}_i^{n},
\dot{\mathbf{q}}_i^{n},
\hat{f}_{i}^{\mathrm{cable},n+1},
a_g^{n+1};
\Theta_{r,i}
\right),
\label{eq:rom_decoder_operator}
\end{equation}
where
\begin{equation}
\Theta_{r,i}
=
\left\{
\mathbf{M}_{r,i},
\mathbf{C}_{r,i},
\mathbf{K}_{r,i},
\mathbf{B}_{i}^{\mathrm{cable}},
\mathbf{B}_{i}^{\mathrm{ground}}
\right\}.
\label{eq:rom_decoder_parameters}
\end{equation}

In this study, $\mathcal{D}_{\mathrm{ROM}}(\cdot)$ is implemented using a matrix form of the Newmark time-integration scheme, so all operations remain differentiable and gradients can be back-propagated from the training loss to the GNN parameters.

The physical response at the top DOF is reconstructed from the predicted modal states as:
\begin{equation}
\hat{u}_{\mathrm{top},i}^{n+1}
=
\mathbf{H}_i
\hat{\mathbf{q}}_i^{n+1},
\quad
\hat{v}_{\mathrm{top},i}^{n+1}
=
\mathbf{H}_i
\hat{\dot{\mathbf{q}}}_i^{n+1},
\quad
\hat{a}_{\mathrm{top},i}^{n+1}
=
\mathbf{H}_i
\hat{\ddot{\mathbf{q}}}_i^{n+1}.
\label{eq:top_response_decoder}
\end{equation}
These reconstructed responses are used in the training loss together with the predicted modal states, as described in the next subsection.

\subsection{Supervised training with modal and response losses}
\label{sec:one_step_training}
The ROMeGNN is trained as a one-step solver using reference data generated from FOM simulations. For each training sample, the FOM response is first projected onto the retained modal basis to obtain the reference modal states $\mathbf{q}_i^n$ and $\dot{\mathbf{q}}_i^n$ at each time step. During training, a time step $n$ is randomly selected, and the graph-state input is constructed using the FOM-projected modal states at time $n$ together with the next-step ground acceleration $a_g^{n+1}$. The model then predicts the next-step modal states and reconstructs the corresponding top responses through the differentiable ROM solver.

The one-step prediction is supervised using both modal-state quantities and reconstructed physical responses. The target quantities include the modal displacement $\mathbf{q}^{n+1}$, modal velocity $\dot{\mathbf{q}}^{n+1}$, top displacement $\mathbf{u}_{\mathrm{top}}^{n+1}$, top velocity $\mathbf{v}_{\mathrm{top}}^{n+1}$, and top acceleration $\mathbf{a}_{\mathrm{top}}^{n+1}$ from the FOM reference solution. The training loss is defined as:
\begin{equation}
\mathcal{L}_{\mathrm{step}}
=
\sum_{\eta \in \{q,\dot{q},u_{\mathrm{top}},v_{\mathrm{top}},a_{\mathrm{top}}\}}
w_{\eta}
\frac{1}{N_{\eta}}
\left\|
\frac{
\hat{\mathbf{y}}_{\eta}^{\,n+1}
-
\mathbf{y}_{\eta}^{\,n+1}
}{
s_{\eta}
}
\right\|_2^2 ,
\label{eq:one_step_loss}
\end{equation}
where $\mathbf{y}_{\eta}^{n+1}$ and $\hat{\mathbf{y}}_{\eta}^{n+1}$ denote the FOM reference and model prediction for quantity $\eta$, respectively. The scale factor $s_{\eta}$ is computed as the standard deviation of the corresponding quantity over the training set so that loss terms with different units and magnitudes can be balanced. $N_{\eta}$ is the number of entries of the corresponding vector, and $w_{\eta}$ is the loss weight.

\subsection{Autoregressive rollout deployment}
\label{sec:Autoregressive rollout deployment}
After training, the ROMeGNN is deployed as an autoregressive forward simulator. Starting from the initial modal state, the model predicts the next-step modal state using the graph topology, component attributes, edge properties, and ground-motion input. The predicted modal states are then recursively fed back as part of the graph-state input for the next time step:
\begin{equation}
\hat{\mathbf{x}}_i^{n+1}
=
\left[
\hat{\mathbf{q}}_i^{n+1},
\hat{\dot{\mathbf{q}}}_i^{n+1},
a_g^{n+2},
\mathbf{s}_i
\right],
\label{eq:rollout_input}
\end{equation}
which enables the model to generate the full time-history response over the duration of a seismic excitation.

During rollout, the learned interaction force $\hat{f}_i^{\mathrm{cable}}$ is directly obtained from the GNN output, while the top responses are reconstructed from the predicted modal states using Eq. \eqref{eq:top_response_decoder}. 
In addition, given that seismic damage to substation equipment is often observed near the root of porcelain bushings or post insulators, the section moment at a predefined critical section is evaluated as an engineering demand quantity for damage assessment. In this study, the moment is recovered from the predicted inertial force distribution by defining a height-weighting vector $\mathbf{h}_i$, whose entries represent the vertical distance between each translational DOF and the selected critical section, so that the moment is approximated by summing the moments of inertial forces about that section.

For the FOM, this moment can be written as:
\begin{equation}
m_i(t)=\mathbf{h}_i^T\mathbf{M}_i\left(\ddot{\mathbf{u}}_i(t)+
\mathbf{r}_i a_g(t)\right),
\label{eq:fom_moment}
\end{equation}
where $\ddot{\mathbf{u}}_i(t)+\mathbf{r}_i a_g(t)$ is the absolute acceleration vector of component $i$. Substituting the modal approximation $\ddot{\mathbf{u}}_i(t)\approx \mathbf{\Phi}_i\ddot{\mathbf{q}}_i(t)$ gives the ROM-based moment recovery equation:
\begin{equation}
\hat{m}_{i}^{n}
=\left(\mathbf{B}_{i}^{\mathrm{moment}}\right)^T
\hat{\ddot{\mathbf{q}}}_{i}^{n}+
G_{i}^{\mathrm{moment}}a_g^{n},
\label{eq:moment_prediction}
\end{equation}
where
\begin{equation}
\mathbf{B}_{i}^{\mathrm{moment}}
=\mathbf{\Phi}_i^T\mathbf{M}_i\mathbf{h}_i,
\quad
G_{i}^{\mathrm{moment}}
=\mathbf{h}_i^T\mathbf{M}_i\mathbf{r}_i .
\label{eq:moment_operators}
\end{equation}
Here, $\mathbf{B}_{i}^{\mathrm{moment}}$ maps the predicted modal acceleration to the section moment, while $G_{i}^{\mathrm{moment}}$ accounts for the contribution of the ground acceleration to the absolute inertial force.

\section{Numerical study}
\subsection{Standalone electrical equipment models}
\begin{figure}[htbp]
    \centering
    \includegraphics[width=0.7\linewidth]{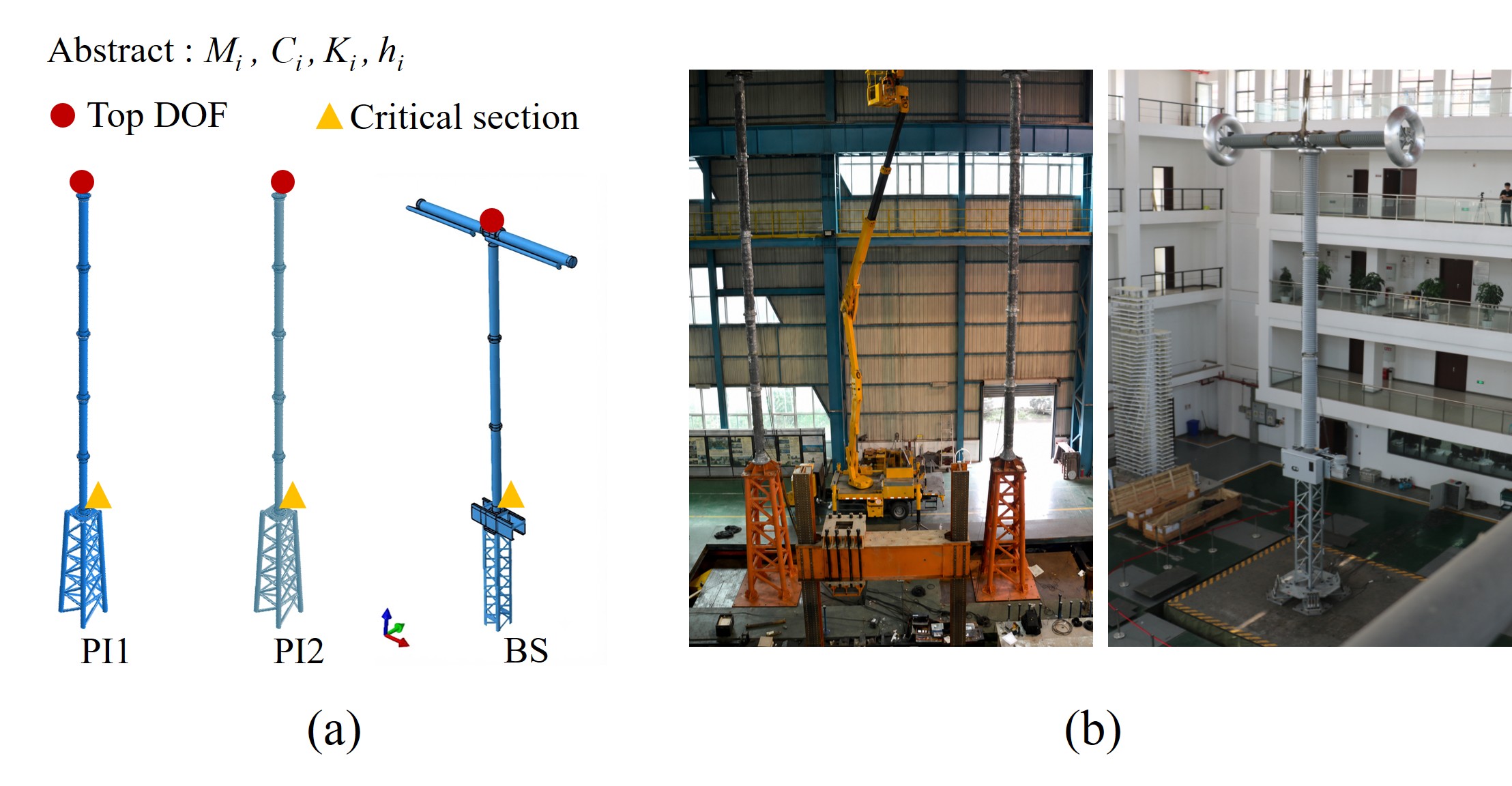}
    \caption{Standalone electrical equipment models used in the numerical study: (a) finite element component library, including two post-insulator components (PI1 and PI2) and one bypass-switch component (BS), with the predefined top DOF and critical section used for response and moment reconstruction; (b) shaking table tests in previous studies \cite{he2022seismic,xie2019seismic} to validate the standalone finite element models.}
    \label{fig:standalone-equipment-models}
\end{figure}

The numerical study uses a component library consisting of three representative types of substation electrical equipment: two post-insulator components, denoted as PI1 and PI2, and one bypass-switch component, denoted as BS, as shown in Figure \ref{fig:standalone-equipment-models}(a). These standalone components serve as the graph nodes, from which different interconnected equipment systems are generated.

The component models are derived from detailed FE models established in ABAQUS in previous studies \cite{he2022seismic,xie2019seismic,liang2025reliability}. The FE models were constructed according to the geometry and structural details of real equipment, and their dynamic characteristics and seismic responses were validated using shaking-table tests of physical specimens, as illustrated in Figure \ref{fig:standalone-equipment-models}(b) \cite{he2022seismic,xie2019seismic}. The corresponding mass, damping, and stiffness matrices are exported from these validated FE models and subsequently reduced to modal ROMs following the procedure described in Section \ref{sec:FEmodal_ROM_construction}.

For each component, a top DOF is defined from a predefined node set in the FE model, where inter-component interaction forces are applied and where the top displacement, velocity, and acceleration responses are recovered. A critical section is also specified for each component to evaluate the section moment associated with seismic demand. The section is defined through a predefined node set in the FE model, and the corresponding nodal coordinate information is used to construct the height-weighting vector $\mathbf{h}_i$ introduced in the previous Section \ref{sec:Autoregressive rollout deployment}. Modal convergence and response reconstruction checks showed that retaining four modes limited the peak-response errors to 0.28\% for top displacement and 0.91\% for section moment across the tested standalone cases. Four modes are therefore retained for each component to construct the modal ROM solver in the study.

\subsection{FOM system assembly and hidden interaction scenarios}
A population of interconnected FOM systems is generated by assembling the standalone component models into line-graph configurations. Adjacent components are connected through mechanical links acting at their top DOFs, and the assembled systems are later used to generate reference responses for training, validation, and testing. For each generated system, the number of components and component types are randomly sampled from the component library, and the nominal connection stiffness and damping coefficients for each physical connection are independently sampled from uniform distributions:
\begin{equation}
k_{c,ij}\sim \mathcal{U}(5.0\times10^3,\,2.0\times10^4),
\quad
c_{c,ij}\sim \mathcal{U}(60,\,300),
\label{eq:connection_sampling}
\end{equation}
with units consistent with the component FE models. In addition to the observable connection parameters, hidden interaction terms are introduced in the reference simulator to emulate unknown connection effects.

\begin{figure}[htbp]
    \centering
    \includegraphics[width=0.9\linewidth]{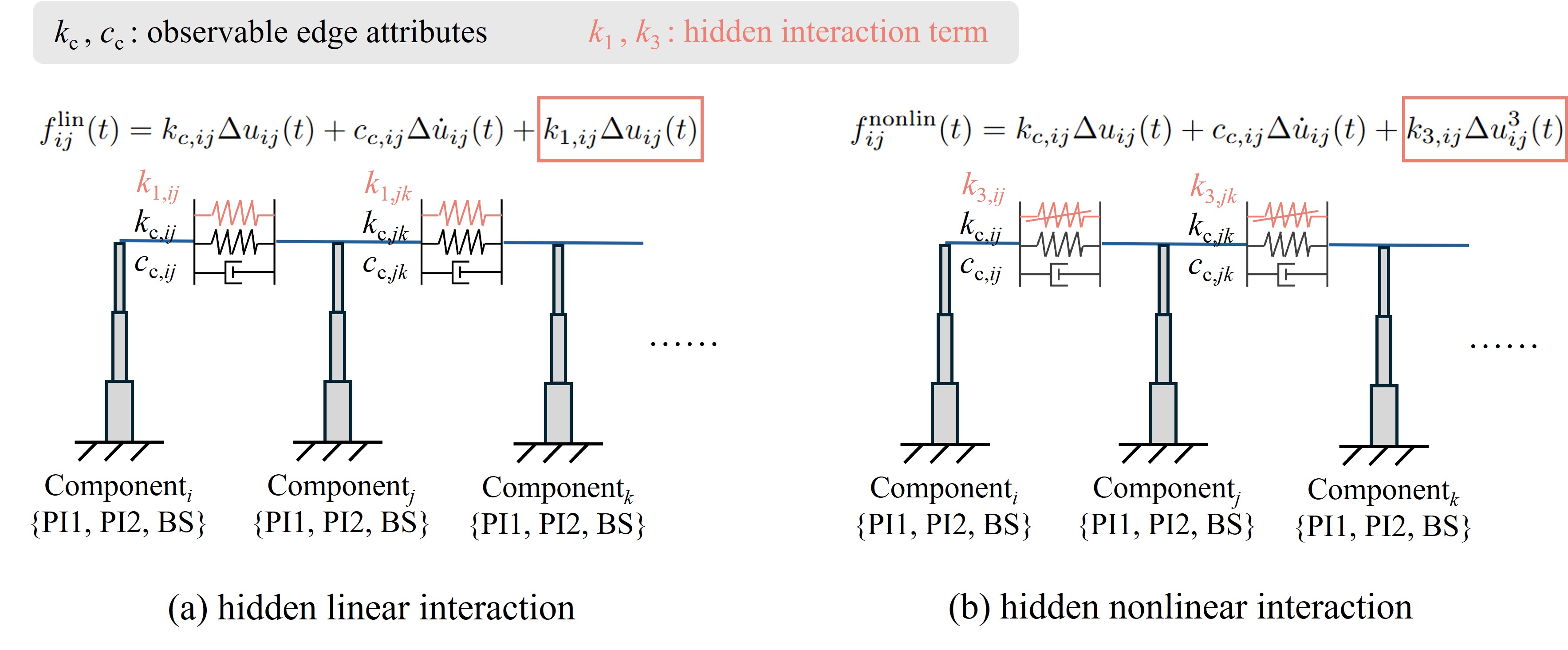}
    \caption{Schematic illustration of the generated line-graph equipment systems and hidden interaction scenarios: (a) hidden linear interaction, where an additional stiffness term $k_{1,ij}\Delta u_{ij}$ is included in the reference simulator; (b) hidden nonlinear interaction, where a cubic stiffness term $k_{3,ij}\Delta u_{ij}^{3}$ is included. The nominal connection stiffness $k_{c,ij}$ and damping $c_{c,ij}$ are provided as observable edge attributes, whereas the hidden terms $k_{1,ij}$ and $k_{3,ij}$ are used only for FOM reference data generation and are not provided to the GNN. Each node type is sampled from the component library $\{\mathrm{PI1},\mathrm{PI2},\mathrm{BS}\}$.}
    \label{fig:hidden-interaction-scenarios}
\end{figure}

Two hidden interaction scenarios are considered to evaluate the ability of the proposed ROMeGNN to learn the unmodeled inter-component dynamics, as illustrated in Figure \ref{fig:hidden-interaction-scenarios}. In the first scenario, a hidden linear stiffness term is introduced into the interaction force:
\begin{equation}
f_{ij}^{\mathrm{lin}}(t)
=k_{c,ij}\Delta u_{ij}(t)+c_{c,ij}\Delta \dot{u}_{ij}(t)+k_{1,ij}\Delta u_{ij}(t),
\label{eq:hidden_linear_interaction}
\end{equation}
where $\Delta u_{ij}(t)=u_{\mathrm{top},j}(t)-u_{\mathrm{top},i}(t)$ and $\Delta \dot{u}_{ij}(t)=v_{\mathrm{top},j}(t)-v_{\mathrm{top},i}(t)$. The hidden stiffness coefficient is set as $k_{1,ij}=0.3k_{c,ij}$, so that the hidden linear stiffness contributes an additional 30\% of the nominal connection stiffness in the reference simulator.

In the second scenario, a hidden nonlinear stiffness term is considered:
\begin{equation}
f_{ij}^{\mathrm{nonlin}}(t)
=
k_{c,ij}\Delta u_{ij}(t)
+
c_{c,ij}\Delta \dot{u}_{ij}(t)
+
k_{3,ij}\Delta u_{ij}^{3}(t),
\label{eq:hidden_nonlinear_interaction}
\end{equation}
where $k_{3,ij}$ controls the cubic stiffness contribution. The cubic stiffness coefficient is set as $k_{3,ij}=0.5k_{c,ij}/u_{\mathrm{ref}}^2$ with $u_{\mathrm{ref}}=0.15$ m, so that the cubic force contribution reaches 50\% of the nominal linear elastic force when $|\Delta u_{ij}|=u_{\mathrm{ref}}$.

The same ROMeGNN architecture and training procedure are used for both scenarios. In both cases, the hidden terms $k_{1,ij}$ and $k_{3,ij}$ are used only in the FOM reference simulator and are not provided to the GNN; their effects must therefore be inferred indirectly through the modal-state and response-level losses.

\subsection{Dataset generation and train-test split}
\label{sec:Dataset_generation}
For each hidden interaction scenario, independent training, validation, and test datasets are generated from the assembled FOM systems described above. For each sample, a synthetic horizontal ground motion acceleration is applied to the assembled system as base excitation. The time-domain FOM response is solved using Newmark integration, and the resulting displacement, velocity, and acceleration histories are used to construct the reference modal states and response quantities described in Section \ref{sec:methodology}.

\begin{figure}[htbp]
    \centering
    \includegraphics[width=0.8\linewidth]{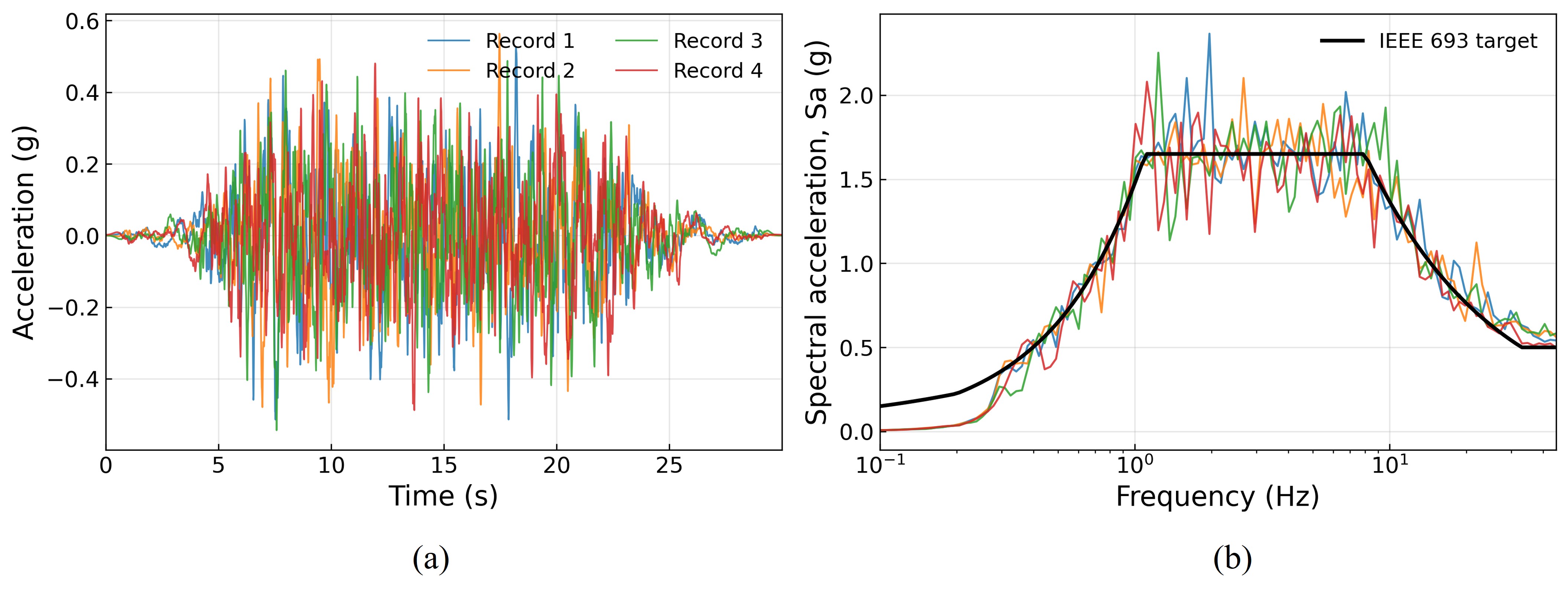}
    \caption{Representative synthetic ground-motion excitations used for dataset generation: (a) acceleration time histories and (b) corresponding response spectra compared with the IEEE 693 target spectrum. The records are first generated to approximately match the target spectral shape and then randomly scaled to provide different excitation intensity levels.}
    \label{fig:ground_motion_generation}
\end{figure}

The ground-motion excitation is generated to be compatible with an IEEE 693-style target response spectrum. Random phase and amplitude variations are introduced, and the acceleration record is adjusted in the frequency domain to approximately match the target spectral shape. To further increase the diversity of seismic demand levels, an overall random scale factor is then applied to the entire acceleration record. As a result, the generated excitations preserve the target spectrum-compatible frequency characteristics while covering different intensity levels. Representative acceleration time histories and their response spectra are shown in Figure \ref{fig:ground_motion_generation}.

The training and validation datasets are generated using graph sizes $N=3$--$8$, where $N$ denotes the number of equipment components in the line-graph system. The test datasets are designed to evaluate both interpolation within the training size range and extrapolation to larger unseen systems. Graph sizes $N=6,7,8$ are used as seen-size but unseen-scenario tests, where the graph sizes are included in the training range but the component combinations, connection properties, hidden parameters, and ground motions are newly sampled. Graph sizes $N=9,10,11,12$ are used as unseen-size tests, where the model is evaluated on larger systems than those encountered during training. Ten independently generated test samples are used for each graph size from $N=6$ to $N=12$, and the prediction errors are reported using the mean and standard deviation across the test samples. The same data-generation and evaluation protocol is used for both interaction scenarios to enable a controlled comparison. The main dataset settings are summarized in Table~\ref{tab:dataset_settings}.

\begin{table}[htbp]
\centering
\caption{Dataset settings used in the numerical study.}
\label{tab:dataset_settings}
\begin{tabular}{lccc}
\hline
Dataset & Graph size $N$ & Number of samples & Purpose \\
\hline
Training & 3--8 & 100 & One-step supervised training \\
Validation & 3--8 & 20 & Validation during training \\
Seen-size test & 6--8 & 10 per size & Unseen scenarios within training size range \\
Unseen-size test & 9--12 & 10 per size & Extrapolation to larger systems \\
\hline
\end{tabular}
\end{table}

\subsection{Implementation details}
\label{sec:implementation_details}
The proposed ROMeGNN is implemented in PyTorch and experiments are conducted on an NVIDIA GeForce RTX 4090 GPU. For the GNN input features, modal states and ground acceleration are normalized by the standard deviations computed from the training set as defined in Section~\ref{sec:one_step_training}. The retained modal frequencies are divided by 10 Hz and included in the static node attributes together with the component type indicator in a one-hot manner. The nominal connection stiffness and damping coefficients are normalized by the upper bounds of their sampling ranges before being used as edge attributes.

Both the edge message MLP and the node update MLP in the message-passing GNN use three layers with a hidden dimension of 128 and SiLU activation functions. The GNN predicts one scalar effective interaction force for each component, which is multiplied by a force scale factor before being passed into the ROM solver. 

The model is trained for 500 epochs with 5000 randomly sampled one-step updates per epoch. The AdamW optimizer is used with a weight decay of $10^{-5}$, and gradient clipping with a maximum norm of 1.0 is applied to improve training stability. A cosine annealing learning-rate scheduler is used, with an initial learning rate of $5\times10^{-4}$ and a minimum learning rate of $10^{-6}$. 
The reported settings were selected based on preliminary hyperparameter screening, and the main parameters are summarized in Table \ref{tab:implementation_settings}.

\begin{table}[htbp]
\centering
\caption{Main implementation settings of the ROMeGNN.}
\label{tab:implementation_settings}
\begin{tabular}{lc}
\hline
Parameter & Value \\
\hline
Retained modes per component & 4 \\
Static node attributes & Component type, retained modal frequencies \\
Observable edge attributes & Normalized $k_{c,ij}$ and $c_{c,ij}$ \\
Hidden dimension & 128 \\
MLP layers & 3 \\
Activation function & SiLU \\
Force scale factor & $1000$ \\
Optimizer & AdamW \\
Weight decay & $10^{-5}$ \\
Initial learning rate & $5\times10^{-4}$ \\
Learning-rate scheduler & Cosine annealing \\
Minimum learning rate & $10^{-6}$ \\
Training epochs & 500 \\
Steps per epoch & 5000 \\
Gradient clipping norm & 1.0 \\
Loss weights & $w_q=w_{\dot q}=w_{u}=w_{v}=1$, $w_a=0.1$ \\
\hline
\end{tabular}
\end{table}

\subsection{Evaluation protocol}
The trained model is evaluated in autoregressive rollout mode. The evaluated quantities include top displacement, top velocity, interaction force, and section moment. The predicted interaction force is obtained from the GNN output and compared with the reference net interaction force computed from the FOM response and the corresponding interaction law used in the reference simulator. The section moment is reconstructed from the predicted modal acceleration using Eq.~\eqref{eq:moment_prediction} and compared with the FOM-based reference moment.

For a response quantity $\mathbf{y}$, the prediction accuracy is evaluated using the relative RMSE defined as:

\begin{equation}
\mathrm{RelRMSE}
=\frac{
\sqrt{
\frac{1}{N_T}
\sum_{n=1}^{N_T}
\left\|
\hat{\mathbf{y}}^{n}
-
\mathbf{y}^{n}
\right\|_2^2}
}{
\sqrt{
\frac{1}{N_T}
\sum_{n=1}^{N_T}
\left\|
\mathbf{y}^{n}
\right\|_2^2}
},
\label{eq:rel_rmse}
\end{equation}

where $\hat{\mathbf{y}}^{n}$ and $\mathbf{y}^{n}$ collect the predicted and reference values of the selected response quantity for all $N$ components in the interconnected system at time step $n$, and $N_T$ is the number of evaluated time steps. RelRMSE is thus computed separately for each test realization using all components and evaluated time steps. For each graph size $N$, the mean and standard deviation of these per-realization values are reported over ten independently generated test cases.

\subsection{Baseline models and sensitivity studies}
\label{sec:Baseline_models}
To assess the contribution of the learned interaction model and the embedded ROM solver, two baseline models are considered. The first baseline is a physics-based coupled-ROM model with incomplete interaction knowledge. In this model, the component ROMs are coupled using only the observable nominal spring-damper parameters $k_{c,ij}$ and $c_{c,ij}$, while the hidden interaction terms $k_{1,ij}$ and $k_{3,ij}$ are excluded. This baseline represents a conventional reduced-order simulation approach in which the standalone component dynamics are known, but the additional inter-component interaction effects are unavailable.

The second baseline is a purely data-driven GNN response predictor. It uses the same graph topology, observable edge attributes, and component static attributes as the proposed model, but does not use modal states, ROM operators, or Newmark-based state advancement. Instead, it directly predicts the next top displacement and velocity in residual form. The section moment is predicted directly as a diagnostic response quantity and is not fed back into the recurrent state. 
To ensure a fair comparison, the pure GNN baseline is configured with a comparable graph-learning capacity to the proposed model: it employs the same message-passing structure, three-layer MLPs with a hidden dimension of 128, and the same activation function. The optimizer, learning-rate schedule, number of epochs, and training datasets are kept the same as those of the proposed ROMeGNN. This baseline is used to examine the benefit of embedding the known structural dynamics into the learning architecture.

In addition, a training-data sensitivity study is conducted to investigate how the amount of FOM reference data affects rollout prediction and generalization performance. For each hidden interaction scenario, a fixed pool of FOM training samples is first generated. Nested training subsets of different sizes are then constructed by taking an increasing number of samples from this pool, and separate models are trained using the same architecture and optimization settings. Since all subsets originate from the same candidate pool, this setup reduces the influence of differences in data-generation conditions and provides a controlled comparison of the effect of training data availability.

\section{Results and discussion}
\subsection{Training behavior}
\begin{figure}[htbp]
    \centering
    \includegraphics[width=1.0\linewidth]{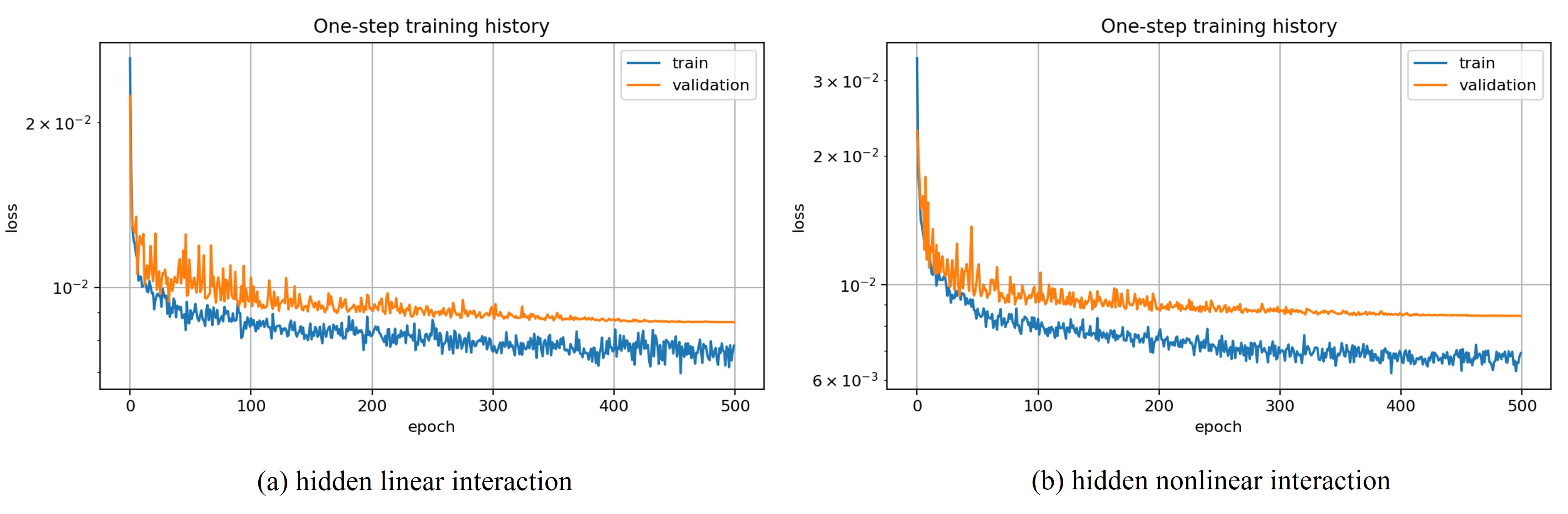}
    \caption{Training histories of the ROMeGNN under (a) hidden linear interaction and (b) hidden nonlinear interaction scenarios.}
    \label{fig:training_history}
\end{figure}

Figure \ref{fig:training_history} shows the one-step training and validation losses of the ROMeGNN under the hidden linear and hidden nonlinear interaction scenarios. Both the training and validation losses decrease consistently and become stable after a few hundred epochs. The validation loss remains slightly higher than the training loss, as expected for unseen validation samples, but the gap between the two curves is small and stable. This indicates that the model generalizes reasonably well at the one-step level and does not exhibit severe overfitting.

It should be noted that the training objective supervises one-step predictions. Therefore, a low one-step loss does not necessarily guarantee accurate long-horizon rollout prediction, where errors may accumulate recursively. The rollout performance under unseen scenarios and unseen graph sizes is evaluated in the following subsections.

\subsection{Rollout prediction on representative unseen systems}
Representative rollout predictions are examined for both hidden linear and hidden nonlinear interaction scenarios. For each scenario, two test systems are selected: a seen-size unseen system with $N=6$ and an unseen-size extrapolation system with $N=10$. For each system, three representative nodes are visualized, and the compared quantities include top displacement, top velocity, section moment, and net interaction force.

\begin{figure}[htbp]
    \centering
    \includegraphics[width=0.93\linewidth]{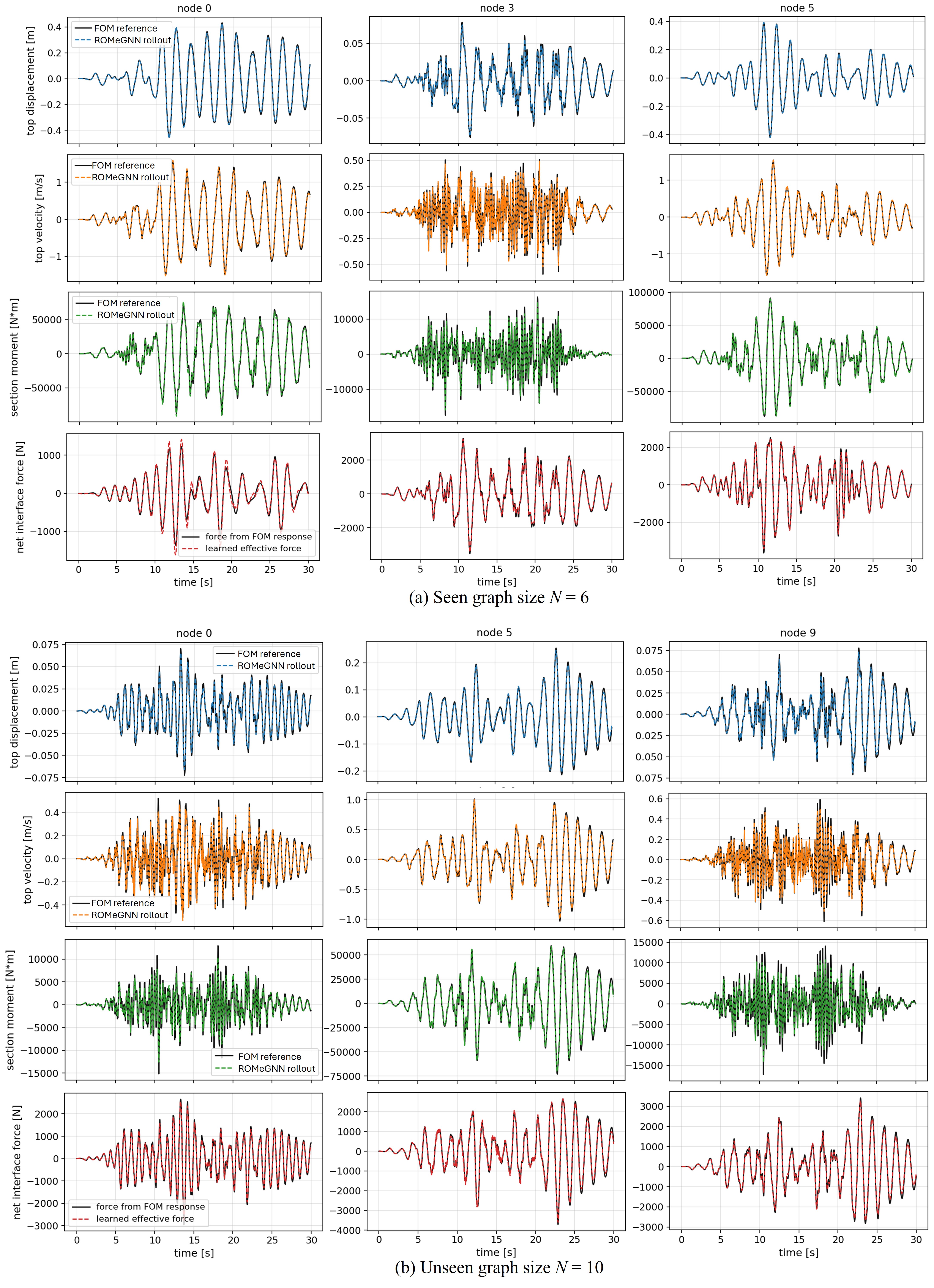}
    \caption{Rollout predictions for the hidden linear interaction scenario: (a) seen-size unseen test system with $N=6$ and (b) unseen-size extrapolation test system with $N=10$. Top displacement, top velocity, section moment, and net interaction force are compared for three representative nodes.}
    \label{fig:linear_rollout_representative}
\end{figure}

Figure~\ref{fig:linear_rollout_representative} shows the results under the hidden linear interaction scenario. For the $N=6$ case in Figure \ref{fig:linear_rollout_representative}(a), the ROMeGNN predictions closely follow the FOM reference responses for all selected nodes. The top displacement and velocity histories are well captured in both phase and amplitude. The section moment predictions also reproduce the dominant peaks and oscillatory patterns, indicating that the predicted modal accelerations provide reasonable engineering demand estimates. The learned effective interaction forces show good agreement with the reference net interaction forces, with only local discrepancies near sharp peaks. For the $N=10$ case in Figure \ref{fig:linear_rollout_representative}(b), the model is evaluated on a larger system than those used during training. Despite this unseen-size extrapolation, the predicted responses remain well aligned with the FOM references across the selected nodes and response quantities. This result suggests that the shared message-passing interaction model and the component-level ROM solver can transfer to larger line-graph systems under hidden linear interaction conditions.

\begin{figure}[htbp]
    \centering
    \includegraphics[width=0.93\linewidth]{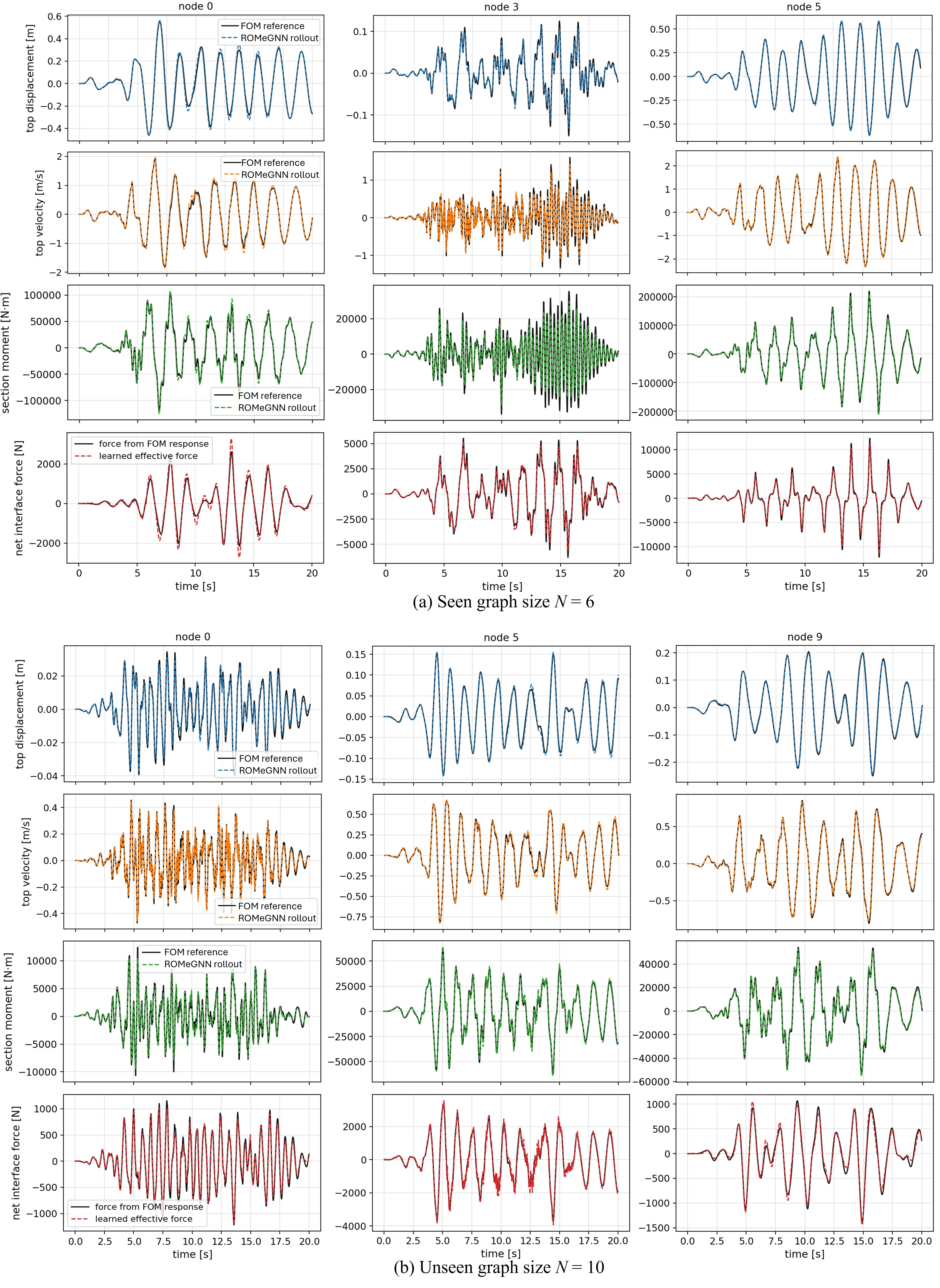}
    \caption{Rollout predictions for the hidden nonlinear interaction scenario: (a) seen-size unseen test system with $N=6$ and (b) unseen-size extrapolation test system with $N=10$. Top displacement, top velocity, section moment, and net interaction force are compared for three representative nodes.}
    \label{fig:nonlinear_rollout_representative}
\end{figure}

Figure \ref{fig:nonlinear_rollout_representative} shows the corresponding rollout predictions under the hidden nonlinear interaction scenario. Compared with the hidden linear case, this scenario is more challenging because the reference simulator contains an unobserved cubic stiffness contribution. Nevertheless, for both the $N=6$ and $N=10$ test systems, the predicted top displacement and velocity responses remain close to the FOM references over the full excitation duration. The section moment histories are also captured with good overall agreement, including the main response peaks and high-frequency oscillatory patterns. The learned effective interaction forces follow the reference net interaction forces reasonably well, although more visible local discrepancies are observed near sharp peaks and highly nonlinear response intervals. These discrepancies are expected because the nonlinear interaction term is hidden from the model input and must be inferred only through the response-level supervision.

Overall, these representative samples indicate that the proposed ROMeGNN can maintain stable autoregressive rollout behavior across both seen-size and unseen-size systems, even when the inter-component interaction contains unknown linear and nonlinear effects. A systematic quantitative assessment over multiple independent test cases and graph sizes is provided in the following subsection.

\subsection{Generalization across graph sizes}
\begin{figure}[htbp]
    \centering
    \includegraphics[width=1\linewidth]{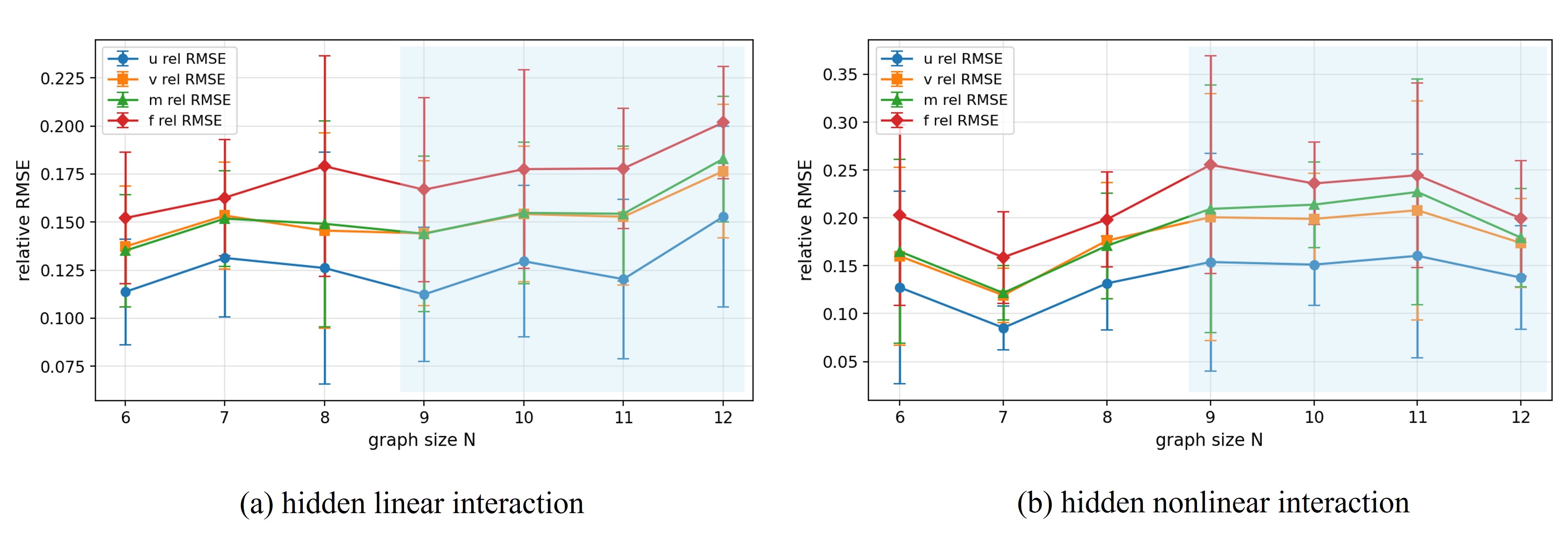}
    \caption{Rollout prediction errors across graph sizes under (a) hidden linear interaction and (b) hidden nonlinear interaction scenarios. Markers indicate the mean relative RMSE, and error bars indicate one standard deviation. The shaded region denotes unseen-size extrapolation cases.}
    \label{fig:error_by_size}
\end{figure}

The generalization performance of the proposed model across graph sizes is evaluated using the test protocol described in section \ref{sec:Dataset_generation}. For compact notation, the relative RMSEs of top displacement, top velocity, section moment, and net interaction force are denoted as $u$, $v$, $m$, and $f$, respectively. The shaded region in Figure \ref{fig:error_by_size} indicates the unseen-size extrapolation cases, where the graph size is larger than those used during training.

Figure \ref{fig:error_by_size}(a) shows the rollout errors for the hidden linear interaction scenario. Across the seen-size test cases with $N=6$--$8$, the displacement error remains lower than the other response quantities, while the velocity and section moment errors show similar levels. This trend is expected because velocity and moment are more sensitive to high-frequency components and local phase shifts than displacement. The net interaction-force error is generally the largest among the four quantities. This is also reasonable because the force is not directly supervised in the training loss and is recovered as an effective interaction input through the response-matching objective.
For the unseen-size cases with $N=9$--$12$, the errors remain in a comparable range to those of the seen-size cases. No monotonic error growth is observed immediately after extrapolating beyond the training size range. This suggests that the learned interaction-force model can be transferred to larger line-graph systems through the shared message-passing architecture, without requiring a fixed graph dimension. The moderate increase at $N=12$ may be related to the stronger extrapolation from the training size range, as well as the longer interaction paths in larger line-graph systems. In such systems, local prediction errors may propagate through more connected components during autoregressive rollout, and the resulting velocity, moment, and interaction-force responses can become more sensitive to small phase or amplitude discrepancies. Part of this variation may also arise from the finite number of randomly generated test cases for each graph size.

Figure~\ref{fig:error_by_size}(b) presents the corresponding results for the hidden nonlinear interaction scenario. Compared with the hidden linear case, the nonlinear case exhibits slightly larger errors and stronger sample-to-sample variability, particularly for the interaction force and section moment. This is expected because the reference simulator contains an unobserved cubic stiffness contribution, while the GNN receives only the nominal linear stiffness and damping as edge attributes. Therefore, the more complicated nonlinear interaction effect must be inferred implicitly from the current graph state and response-level supervision. Despite this additional difficulty, the errors remain bounded across both seen-size and unseen-size cases.

An interesting observation is that the errors in the nonlinear case show a noticeable local increase at $N=9$ and remain relatively high for $N=10$ and $N=11$, before decreasing at $N=12$. This non-monotonic trend indicates that graph size alone does not determine the rollout error; the sampled component types, connection properties, hidden interaction parameters, and ground-motion excitation also affect the prediction difficulty. Overall, the results demonstrate that the proposed ROMeGNN can generalize to larger graph systems and maintain stable rollout behavior under both hidden linear and hidden nonlinear interaction scenarios.

The results were obtained using a fixed evaluation seed for reproducibility. Additional evaluations with different random seeds showed similar overall trends, indicating that the conclusions are not specific to a single test-set realization.

\subsection{Comparison with baseline models}
To evaluate the contribution of the learned interaction-force model and the embedded ROM solver, the proposed method is compared with the baseline models introduced in Section \ref{sec:Baseline_models} with respect to $u$, $v$, and $m$. The interaction force $f$ is not included in this comparison since the pure GNN baseline directly predicts response quantities and does not produce an explicit interaction-force estimate.

\begin{figure}[htbp]
    \centering
    \includegraphics[width=1\linewidth]{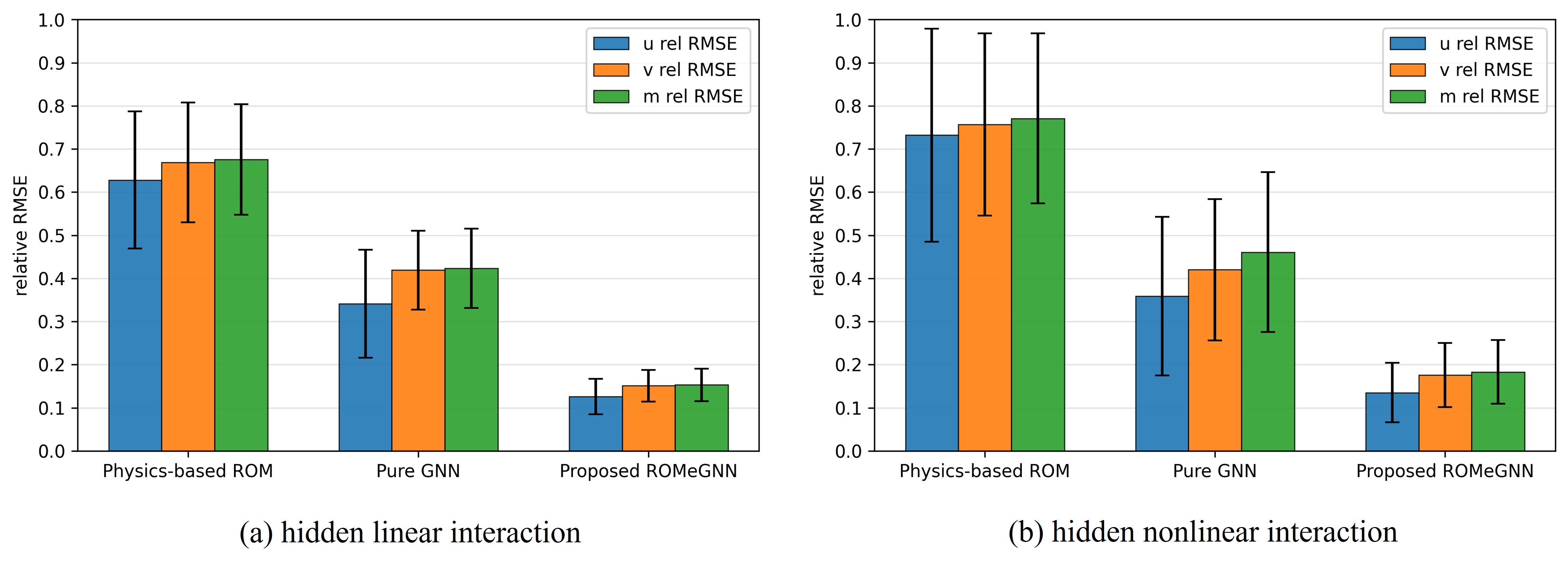}
    \caption{Comparison of rollout prediction errors among the physics-based ROM baseline, pure GNN baseline, and proposed ROMeGNN under (a) hidden linear and (b) hidden nonlinear interaction scenarios. Bars indicate the mean relative RMSE aggregated over test cases, and error bars indicate one standard deviation.}
    \label{fig:baseline_comparison}
\end{figure}

Figure \ref{fig:baseline_comparison} compares the rollout errors over the test graph sizes and independent test cases. For both hidden interaction scenarios, the physics-based ROM baseline gives the largest mean errors and variability. This is expected because it only uses the observable nominal spring-damper connection parameters and therefore cannot represent the hidden interaction terms used in the FOM reference simulator. The effect is more pronounced in the hidden nonlinear case, where the unmodeled cubic interaction leads to larger errors and wider error bars.

The pure GNN baseline reduces the error compared with the physics-based ROM, showing that graph-based learning can partially compensate for the missing interaction physics from data. However, its errors remain substantially larger than those of the proposed method, especially for velocity and section moment. Since the pure GNN directly predicts response quantities without enforcing the component-level equations of motion, it must learn both the standalone structural dynamics and the inter-component interaction effects from data. This can be less effective for long-horizon rollout prediction, particularly for response quantities that are sensitive to phase errors and high-frequency components.

The proposed ROMeGNN achieves the lowest errors in both the hidden linear and hidden nonlinear cases. By embedding the known standalone equipment dynamics in the ROM operator, the GNN does not need to learn the full structural dynamics from data. Instead, it focuses on estimating the effective interaction input between connected components. This division of labor leads to more accurate and more stable rollout predictions than either an incomplete physics-based ROM or a purely data-driven GNN predictor. In addition, the learned interaction force provides a physically interpretable intermediate quantity that is not available from the pure GNN baseline and can be compared with the FOM force references to inspect the inferred hidden interaction dynamics.

Overall, this ablation study demonstrates the complementary roles of the component ROM and the graph-based interaction model. The physics-based ROM baseline highlights the limitation of incomplete physics when hidden interaction effects are present, while the pure GNN baseline highlights the limitation of purely data-driven rollout prediction without embedded structural dynamics. By combining ROM-based state advancement with graph-based interaction learning, the proposed model achieves more accurate and robust predictions across the evaluated response quantities.

\subsection{Training-data sensitivity}
The influence of the number of FOM training scenarios is examined using the sensitivity setting described in Section \ref{sec:Baseline_models}. For this study, the original 100-scenario training sets were expanded to pools of 300 and 150 scenarios for the hidden linear and nonlinear interaction cases, respectively, using the same data-generation procedure. Models are trained on nested subsets of these fixed pool, while the architecture and optimization settings are kept unchanged. The evaluation follows the same test protocol as in the previous subsections.

\begin{figure}[htbp]
    \centering
    \includegraphics[width=1\linewidth]{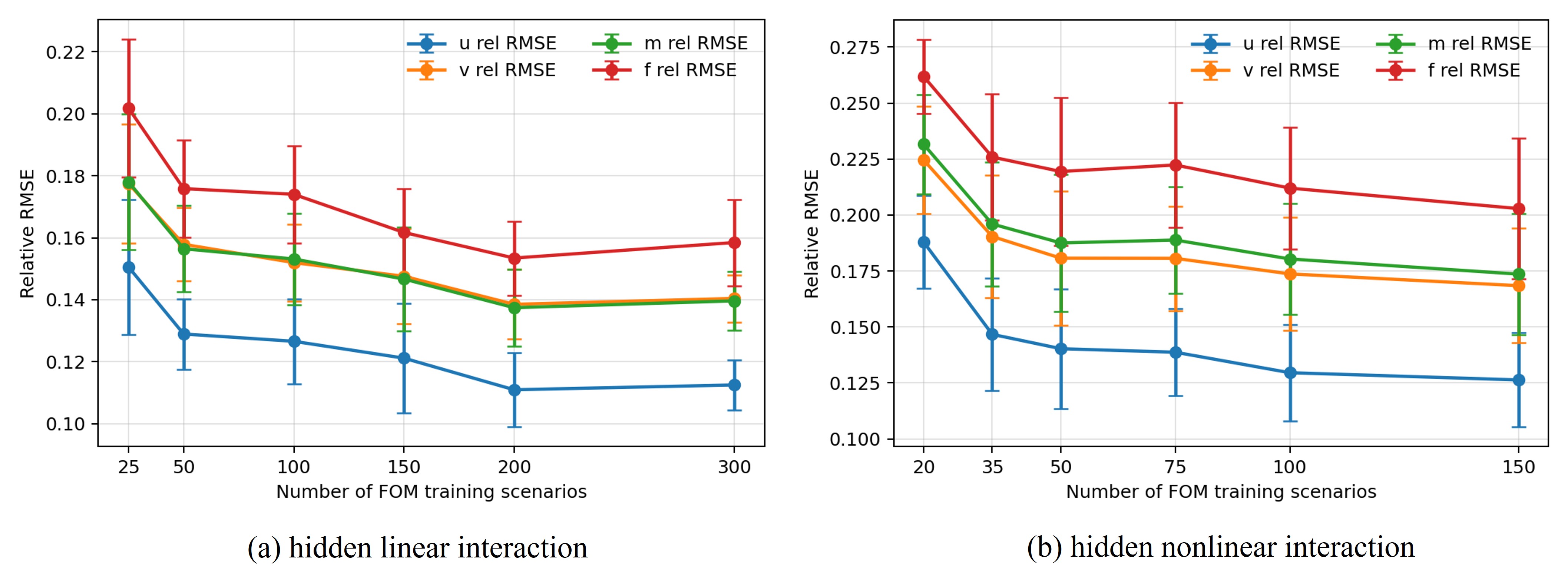}
    \caption{Effect of the number of FOM training scenarios on rollout prediction errors under (a) hidden linear interaction and (b) hidden nonlinear interaction scenarios. Markers indicate the mean relative RMSE, and error bars indicate one standard deviation over independent test cases.}
    \label{fig:training_data_sensitivity}
\end{figure}

Figure \ref{fig:training_data_sensitivity} shows the rollout errors as a function of the number of FOM training scenarios. For the hidden linear interaction scenario, increasing the training set size generally reduces the prediction errors for all response quantities. The improvement is most pronounced when the number of scenarios increases from 25 to 50, indicating that very small training sets may be insufficient to cover the variability in component combinations, connection properties, and excitation characteristics. Beyond approximately 150--200 scenarios, the error reduction becomes more gradual, suggesting that the model begins to approach a saturation regime under the present architecture and training settings.

A similar trend is observed for the hidden nonlinear interaction scenario, although the errors and variability are generally larger, indicating a more challenging learning task. The reduction from 20 to 35 scenarios is substantial, while further increases lead to more moderate improvements. Local non-monotonic variations, such as the increase around 75 scenarios, can arise from stochastic optimization, finite test samples, and the increased diversity of the larger training subsets. Therefore, the curves should be interpreted as statistical trends rather than strictly monotonic learning curves. It should be noted that the apparent earlier saturation in the nonlinear case should not be interpreted as the nonlinear problem requiring less data than the linear problem. The two curves are generated from different training pools and should mainly be read as within-scenario sensitivity trends.

Overall, the results indicate that increasing the number of FOM training scenarios improves the robustness of the ROMeGNN, but the benefit gradually diminishes after a moderate number of scenarios. This suggests that the proposed physics-enhanced formulation can achieve reasonable data efficiency by embedding the known component dynamics and using data primarily to learn the hidden interaction effects.

\subsection{Computational efficiency}
The computational efficiency of the proposed ROMeGNN is evaluated on an interconnected system with 9 components for a 30-s response simulation with 3000 time steps. FOM simulations were run on an Intel Core Ultra 7 258V CPU, whereas ROMeGNN rollouts were performed on the GPU specified in the implementation details section. As summarized in Table~\ref{tab:computational_efficiency}, the FOM simulation requires 144 s for the hidden linear case and 1749 s for the hidden nonlinear case. The much higher cost of the nonlinear case is mainly due to the Newton iterations required to solve the nonlinear interaction force at each time step.

In contrast, the trained ROMeGNN completes the same rollout in approximately 2.64--2.70 s, corresponding to less than 1 ms per time step. This results in speed-ups of approximately $55\times$ and $648\times$ for the hidden linear and hidden nonlinear cases, respectively. The greater acceleration in the nonlinear case highlights the advantage of replacing expensive nonlinear FOM simulations with learned interaction-force prediction and low-dimensional ROM state advancement.

These results demonstrate that, once trained, the proposed model enables rapid rollout predictions suitable for repeated scenario evaluation and near-real-time seismic digital twin applications.

\begin{table}[htbp]
\centering
\caption{Computational efficiency comparison for a 30-s response simulation with 3000 time steps.}
\label{tab:computational_efficiency}
\begin{tabular}{lccccc}
\hline
\multirow{2}{*}{Scenario} 
& \multicolumn{2}{c}{FOM simulation} 
& \multicolumn{2}{c}{ROMeGNN rollout} 
& \multirow{2}{*}{Speed-up} \\
\cline{2-5}
& Total (s) & Step (ms) & Total (s) & Step (ms) & \\
\hline
Hidden linear    & 144  & 48  & 2.64 & 0.88 & $55\times$ \\
Hidden nonlinear & 1749 & 583 & 2.70 & 0.90 & $648\times$ \\
\hline
\end{tabular}
\end{table}

\section{Conclusions}
This study proposed a reduced-order model embedded graph neural network for topology-adaptive seismic response prediction of interconnected electrical equipment systems. The main idea is to embed the known component-level structural dynamics into differentiable modal ROM solvers, while using graph message passing to learn the effective interaction forces between connected equipment. In this way, the neural network is not required to learn the entire structural dynamics from data alone, but instead focuses on the topology-dependent interaction effects that are difficult to model explicitly. This separation between known component dynamics and learned interaction physics reduces the learning burden on the neural network and enables the same model to be applied to systems with different numbers of components and connection layouts.

The proposed framework was evaluated using interconnected systems assembled from finite-element-derived electrical equipment models under spectrum-compatible seismic excitations. Two hidden interaction scenarios were considered, including an additional unobserved linear stiffness term and an unobserved nonlinear cubic stiffness term. The results show that the ROMeGNN can produce stable autoregressive rollout predictions for top displacement, top velocity, section moment, and interaction force. The model maintains consistent performance not only for seen-size unseen test systems, but also for larger unseen-size systems, demonstrating its ability to generalize across graph sizes and component configurations.

The baseline comparison further supports the benefit of combining reduced-order physics with graph-based learning. The physics-based ROM baseline suffers from large errors when the interaction physics is incomplete, while the pure GNN baseline is less accurate because it must learn both standalone structural dynamics and inter-component interactions directly from data. By contrast, the proposed ROMeGNN achieves lower rollout errors in both hidden linear and hidden nonlinear interaction scenarios. In addition, the learned interaction force provides a physically interpretable intermediate quantity for understanding the unknown interaction dynamics.

The training-data sensitivity study indicates that increasing the number of FOM training scenarios generally improves prediction robustness, while the improvement gradually saturates after a moderate number of samples. This suggests that, under the present architecture and data-generation setting, the proposed model can achieve stable rollout performance with a moderate number of FOM reference scenarios. The computational efficiency study demonstrates speed-ups of approximately \(55\times\) and \(648\times\) over the FOM simulations for the linear and nonlinear interaction scenarios, respectively, highlighting the potential for rapid scenario evaluation.

Overall, the proposed framework provides a promising route toward real-time, interpretable, and scalable seismic digital twins for interconnected electrical equipment systems. 
It enables rapid scenario evaluation across varying system sizes, component combinations, connection properties, and excitation inputs, while retaining a direct link to component-level structural dynamics through the ROM solver. Although motivated by interconnected power infrastructures, the proposed framework is generally applicable to interconnected structural systems whose components can be represented by reduced-order models and whose interactions are topology-dependent.

Several limitations should be acknowledged. The present study focuses on line-graph equipment systems and controlled hidden interaction scenarios generated from numerical reference simulators. Future work will extend the framework to more general substation topologies, multi-directional seismic excitations, more complex nonlinear and damage-dependent interaction mechanisms, and validation using experimental or field monitoring data. Further work is also needed to incorporate uncertainty quantification and online model updating for operational digital twin applications.

\section*{Acknowledgments}
The authors would like to acknowledge the funding from the European Union under the program HORIZON-CL5-2023-D4-02-01 for project INBLANC - INdustrialisation of Building Lifecycle data Accumulation, Numeracy and Capitalisation, Grant Number: 101147225.

\bibliographystyle{unsrt}  
\bibliography{refs}

\end{document}